\documentclass[pdflatex,sn-mathphys-num]{sn-jnl}

\usepackage{graphicx}%
\usepackage{multirow}%
\usepackage{amsmath,amssymb,amsfonts}%
\usepackage{amsthm}%
\usepackage{mathrsfs}%
\usepackage[title]{appendix}%
\usepackage{xcolor}%
\usepackage{textcomp}%
\usepackage{manyfoot}%
\usepackage{booktabs}%
\usepackage{algorithm}%
\usepackage{algorithmicx}%
\usepackage{algpseudocode}%
\usepackage{listings}%
\usepackage{makecell}

\theoremstyle{thmstyleone}%
\theoremstyle{thmstyletwo}%

\theoremstyle{thmstylethree}%

\begin{document}

\title[Article Title]{ExSEnt: Extrema-Segmented Entropy Analysis of Time Series}


\author[]{\fnm{Sara} \sur{Kamali}}\email{sara.kamali@uam.es}

\author[]{\fnm{Fabiano}\sur{Baroni}}\email{fabiano.baroni@uam.es}

\author[]{\fnm{Pablo} \sur{Varona}}\email{pablo.varona@uam.es}

\affil[]{\orgdiv{GNB. Dpto. de Ingenieria Informatica.}, \orgname{Autonomous University of Madrid}, \orgaddress{\street{C. Francisco Tomás y Valiente, 11},  \postcode{28049}, \state{Madrid}, \country{Spain}}}


\abstract{We introduce \emph{Extrema-Segmented Entropy} (ExSEnt), a feature-decomposed framework for quantifying time-series complexity that separates temporal from amplitude contributions. The method partitions a signal into monotonic segments by detecting sign changes in the first-order increments. For each segment, it extracts the interval duration and the net amplitude change, generating two sequences that reflect timing and magnitude variability, respectively. Complexity is then quantified by computing sample entropy on durations and amplitudes, together with their joint entropy. This decomposition reveals whether overall irregularity is driven by timing, amplitude, or their coupling, providing a richer and more interpretable characterization than unidimensional metrics. We validate ExSEnt on canonical nonlinear dynamical systems (Logistic map, R\"ossler system, Rulkov map), demonstrating its ability to track complexity changes across control parameter sweeps and detect transitions between periodic and chaotic regimes. Then, we illustrate the empirical utility of ExSEnt metrics to isolate feature-specific sources of complexity in real data (electromyography and ankle acceleration in Parkinson’s disease). ExSEnt thus complements existing entropy measures by attributing complexity to distinct signal features, improving interpretability and supporting applications in a broad range of domains, including physiology, finance, and geoscience.
}




\maketitle

\section{Introduction}
\label{sec: Introduction}

\noindent The study of time-series complexity is central to understanding the behavior and predictability of nonlinear dynamical systems. Many natural and engineered processes---from physiological rhythms to markets and climate---exhibit intricate temporal organization that demands robust quantitative descriptors \cite{henry96,bandt2002pe,tang2015complexity,zunino2017detecting,lau2022brain}. Entropy-based metrics have become standard tools to quantify unpredictability, disorder, and structure in temporal data \cite{shannon1948mathematical,pincus1991approximate,lake2002sample,costa2002multiscale,fadlallah2013weighted,rosso2001wavelet,zanin2012permutation,Amigo24}, with extensive applications in physiological signal analysis \cite{richman2000physiological, azami2023entropy, yu2024periodic, costa2005multiscale, fusheng2001approximate}, finance \cite{wang2022efficient, wang2021covid}, climate \cite{silva2021complexity}, and network dynamics \cite{kulisiewicz2018entropy, lei2019improved}. Foundational work links macroscopic complexity to healthy adaptive function, with characteristic loss of complexity in aging and disease \cite{lipsitz1992jama,goldberger2002pnas}, motivating precise and interpretable complexity quantification.

Classical measures such as Shannon entropy, Approximate Entropy, Sample Entropy (SampEn), and Permutation Entropy (PE) probe different aspects of irregularity \cite{shannon1948mathematical, pincus1991approximate, richman2004sample, bandt2002pe}. Multiscale Entropy (MSE) refines this by assessing irregularity across scales \cite{costa2002multiscale,costa2005multiscale}, with algorithmic improvements including refined composite MSE for short/noisy signals \cite{wu2014rcmse}. Symbolic approaches like PE are attractive for noise-robustness and computational efficiency, yet in their basic form, they discard amplitude information \cite{bandt2002pe,riedl2013pe_tutorial}. Several extensions explicitly reincorporate amplitude or fluctuation content, e.g., Weighted-PE \cite{fadlallah2013weighted}, Amplitude-Aware PE \cite{azami2016aape}, Increment Entropy \cite{liu2016increment}, and Multiscale Increment Entropy \cite{wang2022multiscale_increment}. WPE modifies the computation of PE by weighting the occurrence of each ordinal pattern according to the amplitude (local variance or energy) of the values forming it. In contrast, AAPE extends PE by directly incorporating the relative magnitudes of neighboring samples into the symbolization step, thereby embedding amplitude information into the pattern definition itself. Another method that takes into account the amplitude variations is the slope entropy, which extends permutation-based approaches by encoding subsequences according to the slope between consecutive data samples. Symbols are assigned based on thresholded differences (positive, negative, or near-zero), thereby incorporating both the sign and magnitude of local changes into the entropy calculation \cite{cuesta2019slope}. Dispersion Entropy and its multiscale and refined variants further improve stability and speed for biomedical data \cite{rostaghi2016disen,azami2017rcmde}. Practical guidance on PE-family parameterization and pitfalls is now well documented \cite{riedl2013pe_tutorial,zunino2017ties,little2017vectorpe,delgadobonal2019tutorial,timme2018tutorial}.

Despite broad uptake, a key limitation persists: conventional entropy measures rarely disentangle distinct sources of complexity within a single signal. In most real systems, irregularity arises from: \emph{(i)} variability in the timing of upward/downward drifts (durations of such drifts), and \emph{(ii)} variability in the magnitudes of those drifts (amplitude changes). Standard estimators typically conflate these contributions. For example, ordinal analyses emphasize rank order (timing/order relations) but not magnitudes unless explicitly modified (AAPE, WPE) \cite{bandt2002pe,Arroyo2013,azami2016aape, fadlallah2013weighted}, and many MSE implementations inherit this ambiguity \cite{costa2002multiscale,wu2014rcmse,kosciessa2020mse,grandy2016sparse}. Although multivariate and bivariate entropies and mutual-information frameworks capture dependencies across channels or features \cite{rossi1992non,rajesh2014bivariate,kundu2017bivariate,journel1993entropy,ahmed2011multivariate,harder2013redundant,kraskov2004mi}, they do not provide a feature-level decomposition of a single time series into temporal- vs. amplitude-driven components of complexity. The need for such decomposition is underscored in contexts where timing (phase/duration) governs behavior more than amplitude (e.g., specific oscillatory neural dynamics): signal-processing theory formally separates instantaneous phase and amplitude via analytic signals and the Hilbert transform \cite{boashash1992procIEEE,picinbono1997tsp}, but an entropy formulation that quantifies their independent contributions has been missing.

Such decomposition provides insight and understanding in multiple contexts. For example, in neural time series, temporal variations relate to oscillatory dynamics, synchrony, and phase relations; amplitude variations reflect power/gain changes and can be disproportionately affected by artifacts or non-neural influences, particularly when identifying functional variability. Conflation reduces the interpretability of EEG complexity biomarkers. Studies comparing entropy indices in anesthesia and clinical EEG highlight sensitivity to algorithmic choices and to different noise/artifact regimes \cite{liang2015anesthesia,aho2015bja,su2016mspe,ra2021scirep,wu2020applsci,sklenarova2023scirep}, and emphasize the importance of scale-aware, robust, and interpretable metrics \cite{kosciessa2020mse,grandy2016sparse,delgadobonal2019tutorial}. A decomposition that attributes overall complexity to timing versus amplitude could sharpen inference, improve predictive performance, and clarify mechanistic hypotheses.

To address these limitations, we introduce \emph{Extrema-Segmented Entropy (ExSEnt)}, a feature-driven, noise-robust framework that segments a signal by the sign of its increments, forming monotone segments between extrema. For each segment, we extract duration $\mathcal{D}$ and net amplitude change $\mathcal{A}$, and then compute SampEn on the $\mathcal{D}$ and $\mathcal{A}$ sequences separately ($\mathcal{H}_{\mathcal{D}}$, $\mathcal{H}_{\mathcal{A}}$) alongside their joint entropy $\mathcal{H}_{\mathcal{DA}}$. This yields an interpretable decomposition, revealing whether complexity arises primarily from temporal irregularity, amplitude variability, or their interaction. If $\mathcal{H}_{\mathcal{DA}}<\min(\mathcal{H}_{\mathcal{D}},\mathcal{H}_{\mathcal{A}})$, there is some level of coherency in the variations of $\mathcal{D}$ and $\mathcal{A}$, which makes their paired sequence more predictable than when considered alone; if $\mathcal{H}_{\mathcal{DA}}>\max(\mathcal{H}_{\mathcal{D}},\mathcal{H}_{\mathcal{A}})$, the features contribute independently to the complexity of the signal with lower predictability of their paired sequence.

We validate ExSEnt on canonical chaotic systems---the Logistic map, the R\"ossler system, and the Rulkov map~\cite{logistic1976simple, rossler1976equation, rulkov2002modeling}---which undergo well-charac\-terized transitions between periodic and chaotic regimes. Unlike single-number metrics, ExSEnt tracks $\mathcal{H}_{\mathcal D}$, $\mathcal{H}_{\mathcal A}$, and $\mathcal{H}_{\mathcal{DA}}$ along control parameter sweeps and exposes whether complexity changes are driven by temporal irregularity, amplitude variability, or their coupling, thereby enhancing interpretability. Beyond nonlinear systems analysis, this feature-level decomposition is directly relevant for EEG/MEG complexity biomarkers in epilepsy, anesthesia, sleep, and disorders of consciousness, and for biometric and forensic signal analysis where the source of complexity affects robustness and specificity. To illustrate these benefits in real-world biosignals, we evaluated complexity in electromyography (EMG) at rest versus movement, and in lower limb motion acceleration of individuals with Parkinson's disease (PD), to demonstrate how ExSEnt disentangles temporal- versus amplitude-driven contributions.

The remainder is organized as follows. Section~\ref{sec:methodology} details ExSEnt's formulation and implementation. Section~\ref{sec:results} applies ExSEnt to chaotic benchmarks and real physiological data, demonstrating sensitivity to regime transitions and the ability to attribute complexity changes to timing and/or amplitude. Section~\ref{sec:conclusion} discusses broader implications and applications.

\section{Methodology}
\label{sec:methodology}

\noindent We propose a novel method, \textit{ExSEnt}, to quantify the complexity of a signal by leveraging its local bivariate variations. The core idea is to segment the signal based on local extrema, extract duration and amplitude features from each segment, and then apply entropy measures to evaluate their complexity. The methodology consists of three steps:

\noindent\underline{\textit{Segmentation based on local extrema:}} The local minima and maxima of the signal are extracted. Every interval between two consecutive local extrema is considered as a segment.

\begin{figure}[ht!] 
    \centering
    \includegraphics[width=.8\textwidth]{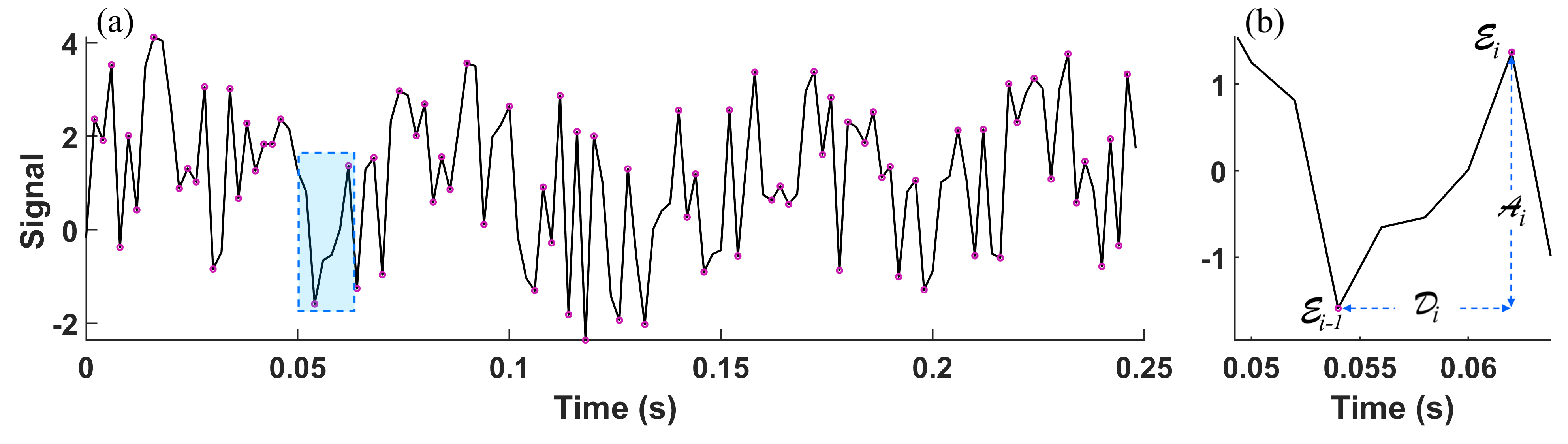}
    \vspace*{-2mm}
    \caption{Sample signal with extracted features for one segment included between consecutive local extrema, $\mathcal{E}_{i-1}$ and $\mathcal{E}_i$. Panel (a) shows an illustrative signal. Panel (b) shows the zoomed plot of the highlighted box from panel (a), with marked extrema and the corresponding extracted features, i.e., the duration and amplitude of that segment.}
    \label{fig:sample_series}
\end{figure}

\noindent\underline{\textit{Feature extraction:}} Two key features, i.e., the duration and the net amplitude change, are extracted for each segment. Fig. \ref{fig:sample_series} depicts this step for one segment.

\noindent\underline{\textit{Entropy computation:}} SampEn is computed on the extracted features and their paired sequence.

\noindent Each of these steps is described in detail in the subsequent sections.

\subsection{Local extrema-based segmentation}
\label{subsec: Segment extraction based on increment sign}

\noindent  Let \( X={x_1,x_2,...,x_N} \) denote the time series of interest. To detect the local extrema, we define the first-order amplitude increment for the $k^{th}$ sample as:
\begin{equation}
    \Delta x_k = x(t_k) - x(t_{k-1})\,.
\end{equation}
A segment is defined as a continuous set of data points, during which the sign of all $\Delta x_k$ values remains constant. Hence, the local extremum, $\mathcal{E}_i$, is defined as the point where the sign of $\Delta x_k$  is not the same as the amplitude increment of its following sample, i.e., $\Delta x_{k+1}$. To enhance robustness against noise, a noise threshold $\theta$ is applied, set to $\lambda$ times the interquartile range (IQR) of the set of all increments of the time series, i.e., $\{\Delta x_k\}$:


\begin{equation}
    \theta=\lambda \times \text{IQR}(\{\Delta x_k\})\,.
\end{equation}

\noindent Hence, an increment $\Delta x_{k+1}$ marks the beginning of a new segment only if:

\begin{equation}
\label{eq:theta_extremum}
|\Delta x_{k+1}|>\theta \;\text{ and }\; \Delta x_k\,\Delta x_{k+1}<0\,.
\end{equation}

\noindent Thus, the noise tolerance depends on $\lambda$. A small $\lambda$ captures finer fluctuations, whereas a larger value suppresses noise-induced variations. In our validation experiments, we set $\lambda = 0.01$, but this parameter may be tuned depending on the dataset and noise characteristics.

\subsection{Feature Extraction}
\label{subsec: Feature Extraction}
 
\noindent After segmentation, we extract two fundamental features per segment: first, the segment duration ($\mathcal{D}$), i.e., the interval between the two extreme points, $\mathcal{E}_i$ and $\mathcal{E}_{i-1}$:
\begin{equation}
    \mathcal{D}_i =  t_{\mathcal{E}_i}-t_{\mathcal{E}_{i-1}}\,;
\end{equation}
second, the amplitude ($\mathcal{A}$) of the segment, i.e., the net amplitude change between the consecutive local extrema:
\begin{equation}
    \mathcal{A}_i = x(t_{\mathcal{E}_i}) - x(t_{\mathcal{E}_{i-1}})\,,
\end{equation}
where the sign of the magnitude differences is preserved.

These features form two separate time series, $\{\mathcal{D}_i\}$ and $\{\mathcal{A}_i\}$, which are the basic elements for computing the ExSEnt metrics. As illustrated in Fig. \ref{fig:sample_series}, our goal is to study the pattern of local extrema-based segment duration and amplitude by analyzing the sequence of extracted $\{\mathcal{D}_i\}$ and $\{\mathcal{A}_i\}$ for $i=1, \dots, M$.

\subsection{Entropy of extracted features}
\label{subsec: entropy of extracted features}

\noindent We selected SampEn to quantify the irregularity of the extracted features because it is particularly advantageous for small datasets. It measures the probability that two sequences of length $m$, similar within a tolerance $r$, remain similar when extended to length $m+1$. The mathematical definition is:
\begin{equation}
    \operatorname{SampEn}(m, r, N) = -\log \left( \frac{\sum_{i=1}^{N-m-1} A_i(m,r)}{\sum_{i=1}^{N-m} B_i(m,r)} \right) = -\log \left( \frac{A}{B} \right)\,,
    \label{eq: sampen}
\end{equation}
\noindent where $m$ is the embedding dimension (length) of the sequences and $r$ is the tolerance threshold between two sequences to measure similarity, which is typically set as a fraction (usually $20\%$) of the standard deviation of the time series. $N$ is the total number of data points. SampEn computation is based on the frequency of appearance of templates. A template of length $m$, $\mathbf{T}^{(m)}$, refers to $m$ consecutive data points in the series. Two templates are considered a match if their distance is within the threshold $r$. To compute this, different measures such as Euclidean distance or Chebyshev distance can be used; the latter is more common:

\begin{equation}
d(\mathbf{T}_i^{(m)}, \mathbf{T}_j^{(m)}) 
= \max_{k=1,\ldots,m} \left| T_i^{(m)}(k) - T_j^{(m)}(k) \right| \leq r\,.
\label{eq:cheby}
\end{equation}

Hence, $B_i$ is the number of matches for all the template vectors of length $m$, i.e,, the number of all indices $j\neq i$ where the difference between $\mathbf{T}_{j}^{(m)}$  and $\mathbf{T}_{i}^{(m)}$ is less than a given threshold, $r$, where $\mathbf{T}_{i}^{(m)}$ is the template of length $m$, starting at sample $i$. Similarly, $A_i$ is the number of matches for the template vector of length $m+1$, starting at the position $i$, except for the template at position $i$ itself. 

\begin{equation}
B = \sum_{i=1}^{M-m} \sum_{j=i+1}^{M-m+1} \mathbb{I} \left( d(\mathbf{T}_i^{(m)}, \mathbf{T}_j^{(m)}) \leq r \right),
\label{eq:sampen B}
\end{equation}
where $\mathbb{I}(\cdot)$ is the indicator function, which returns $1$ if the condition is met and $0$ otherwise.

Similarly, the number of matched pairs for templates of length  $m+1$ is computed as:

\begin{equation}
A = \sum_{i=1}^{M-m-1} \sum_{j=i+1}^{M-m} \mathbb{I} \left( d(\mathbf{T}_i^{(m+1)}, \mathbf{T}_j^{(m+1)}) \leq r \right).
\label{eq:sampen A}
\end{equation}
The set of all possible templates is constructed by taking all the windows of length $m$ (or $m+1$) from the time series and computing the matching pairs. 

The idea of the sample entropy is based on finding the probability that a matching template of length $m$ remains a match when extended to length $m+1$. A lower SampEn value indicates a more predictable signal, while a higher value suggests increased complexity or randomness.

\subsection{Entropy of extracted features}
\label{subsec: Entropy computation on extracted features}

\noindent To quantify the complexity of the signal with the ExSEnt method, we compute the sample entropy of the sequence of extracted features: the durations of the intervals between each pair of consecutive extrema, $\{\mathcal{D}_k\}$, their net amplitude differences, $\{ \mathcal{A}_k\}$, and the paired sequence of these features, $\{(\mathcal{D}_k,  \mathcal{A}_k)\}$, as follows:  

\begin{equation}
    \mathcal{H}_\mathcal{D} = \operatorname{SampEn}(m, r, \{ \mathcal{D}_k\})\,,
\end{equation}

\begin{equation}
    \mathcal{H} _\mathcal{A} = \operatorname{SampEn}(m, r, \{ \mathcal{A}_k\})\,,
\end{equation}

\begin{equation}
    \mathcal{H}_{\mathcal{DA}} = \operatorname{SampEn}(m, r, \{( \mathcal{D}k,  \mathcal{A}_k)\})\,.
\end{equation}
These three entropy measures offer complementary perspectives on the complexity of the signal, capturing different aspects of its underlying dynamics. By analyzing these measures collectively, a more comprehensive characterization of the signal's dynamics is achieved, providing deeper insights into its intrinsic nature.

$\mathcal{H}_\mathcal{D}$ quantifies the degree of temporal variability of segments, capturing whether it follows an underlying pattern or exhibits irregular behavior. Similarly, $\mathcal{H}_\mathcal{A}$ measures the diversity of segments' amplitude variations, reflecting whether the trend of amplitude fluctuations is recurrent or more erratic. 

Meanwhile, the joint entropy $\mathcal{H}_{\mathcal{DA}}$ characterizes the relation between these two features, assessing the extent to which these two sequences are independent. A higher $\mathcal{H}_{\mathcal{DA}}$ suggests a weaker coupling between duration and amplitude changes, indicating more independent or complex interactions between these features. Conversely, lower values suggests stronger synchronization, pointing to structured dependencies in the signal’s temporal and amplitude modulations.

\subsection{Computation of the joint entropy}
\label{subsec: Computation of the joint entropy}

\noindent To compute the joint entropy, first we construct an $M\times2$ matrix, where each row consists of a segment's duration and amplitude pair, and $M$ is the number of extracted segments. At each step $i$, a sequence of $m$ consecutive pairs is extracted and combined into a single vector, forming the $i^{th}$ template:
\[
\mathbf{P}_i^{(m)} = \left( \mathbf{p}_i,~\mathbf{p}_{i+1},~\dots,~\mathbf{p}_{i+m-1} \right)\,,
\]
where $\mathbf{p}_i =(\mathcal{D}_i,\mathcal{A}_i)$ is a row vector, representing the paired duration and amplitude of the $i^{th}$ segment. The concatenation of these row vectors for all the $M$ segments results in the template matrix, $\mathbf{P}^{(m)}$, with dimension $(M-m+1) \times 2m$:

\[
\mathbf{P}^{(m)} = \left[ \mathbf{p}_1^{(m)}, \mathbf{p}_2^{(m)}, \dots, \mathbf{p}_{(M-m+1)}^{(m)}\right]\,.
\]
Similarly, templates of length \( m+1 \) are constructed as:

\[
\mathbf{p}_i^{(m+1)} = \left( \mathbf{p}_i, \mathbf{p}_{i+1}, \dots, \mathbf{p}_{i+m} \right)\,.
\]
\noindent which yield the template matrix of dimension $(M-m) \times 2(m+1)$:

\[
\mathbf{P}^{(m+1)} = \left[ \mathbf{p}_1^{(m+1)}, \mathbf{p}_2^{(m+1)}, \dots, \mathbf{p}_{(M-m)}^{(m+1)}\right]\,.
\]
The similarity between the two templates is determined based on their Chebyshev distance (Eq. \ref{eq:cheby}).

The nature of the paired data, where the $\mathcal{D}$ and $\mathcal{A}$ sequences are generated from two different scales and units, typically requires an adjustment. So, before pairing these two sequences, we normalize them to zero mean and unit variance to achieve comparable measures.

For each pair of templates \( (\mathbf{P}_i^{(m)}, \mathbf{P}_j^{(m)}) \), the number of matched pairs is computed as Eqs. (\ref{eq:sampen A}, \ref{eq:sampen B}). The sample entropy is thus calculated as in Eq. (\ref{eq: sampen}):

\[
\mathcal{H}_{\mathcal{DA}} = -\log \left(\frac{A}{B} \right).
\]

To compute the joint sample entropy of the paired set $(\mathcal{D}_i,\mathcal{A}_i)$, it is crucial to maintain consistency with the individual sample entropy evaluations of $\mathcal{D}_i$ and $\mathcal{A}_i$. By preserving the same $m$, the complexity of the joint system is assessed on the same scale as its individual components, making the comparison of results possible and meaningful.

The joint entropy provides insight into the mutual information between segment durations and amplitudes. If $\mathcal{H}_{\mathcal{DA}}$ is greater than or equal to the entropy of either durations or amplitudes, it suggests lower mutual information, implying that durations and amplitudes contain largely independent information. Conversely, if it is lower than both individual entropies, this suggests statistical dependence and coupling between the two variables, as some variability in one metric is already reflected in the other. This methodology allows for the quantification of the degree of temporal and amplitude-driven coupling between local extrema-based segments within the time series, helping to better characterize the system’s dynamical structure.

\section{Results}
\label{sec:results}

\noindent In this section, we present the results obtained by applying the ExSEnt method to synthetic and real benchmark data from various origins: stochastic signals, widely studied chaotic systems---the Logistic Map (LM), the Rössler System (RS), and the Rulkov map (RM)---and two human movement datasets---electromyography and inertial time series. The results focus on the effect of data length on entropy measures and the analysis of metrics over bifurcation diagrams. Also, a comparison of metrics for human data in different states is provided.


\subsection{ExSEnt for representative stochastic signals}
\label{subsec: ExSEnt for ExSEnt for stochastic and periodic data}

\begin{figure}[ht!]
    \centering
    \includegraphics[width=0.9\textwidth]{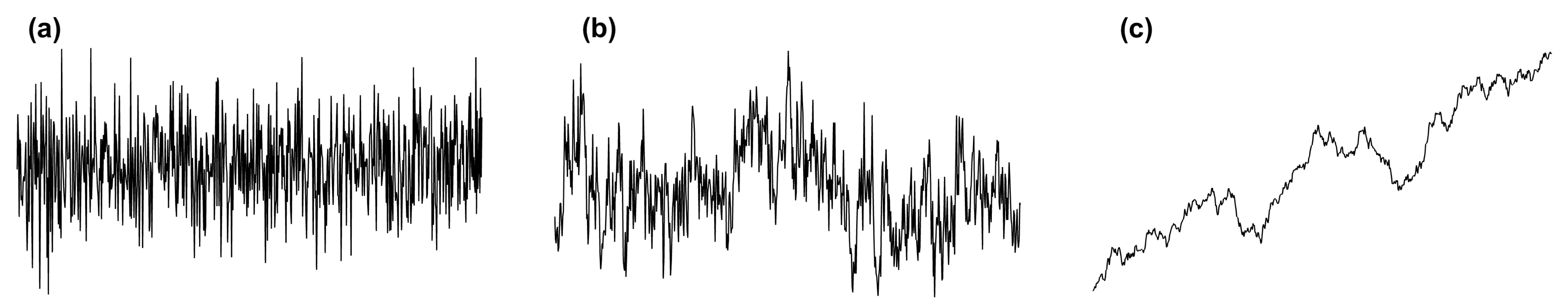}
    \vspace*{.5mm}
    \caption{Representative stochastic signals. Panels (a) to (c) show Gaussian noise, Pink noise, and the Brownian motion, respectively. The signals are plotted over $N=700$ samples, according to the formula presented in section \ref{math sec: stochastic}} 
    \label{Fig:stochastic sig} 
\end{figure}

\noindent  We computed the ExSEnt metrics for Gaussian noise, Pink noise, and Brownian motion (Fig. \ref{Fig:stochastic sig}). Table~\ref{tab:exsent_results} summarizes the statistics of the ExSEnt measures computed over $100$ runs, along with the SampEn.

The results revealed distinct patterns across the signals. For Gaussian noise, the amplitude entropy $\mathcal{H}_{\mathcal{A}}$ was higher than the other ExSEnt metrics, reflecting substantial randomness in the signal's amplitude fluctuations, while the comparably lower $\mathcal{H}_{\mathcal{D}}$ suggested less variability in segment durations.
Pink noise exhibited both higher duration and amplitude entropies compared to Gaussian noise, and its resulting joint entropy was slightly higher than that of Gaussian noise. In the case of Brownian motion, it showed the highest duration entropy compared to the other signals, demonstrating the large variability in the lengths of the segments, while its amplitude entropy remained marginally lower than Pink noise.

Comparison of these numbers with the SampEn results demonstrated the extended ability of ExSEnt metrics to quantify the complexity of the signal. As for both Gaussian and Pink noise, SampEn produces higher entropies, while the calculated entropy for the Brownian is considerably small with a relatively high CV, not reflecting its complex dynamics. In all the analyses, the embedding dimension $m$ was set to 2, unless stated otherwise. The $r$ for computation of sample entropy of sequences is typically set as a fraction (usually 20\%) of the standard deviation of the time series, and we used this ratio consistently with most of the SampEn literature.

\begin{table}[ht]
  \centering
  \caption{ExSEnt vs. SampEn Measures: $N=20000$, $\lambda=0.01$, $m=2$. 
  Values are mean $\pm$ SD along with the CV (second line of each row); 100 seeds. Refer to \ref{math sec: stochastic} for the detailed formula.}
  \label{tab:exsent_results}

  \begingroup
  \fontsize{7.5}{9}\selectfont   
  \renewcommand{\arraystretch}{2} 

  \begin{tabular}{lcccc}
    \toprule
    \textbf{Signal} & $\boldsymbol{\mathcal{H}_{\mathcal D}}$ & $\boldsymbol{\mathcal{H}_{\mathcal A}}$
                    & $\boldsymbol{\mathcal{H}_{\mathcal{DA}}}$ & \textbf{SampEn} \\
    \midrule
    Gaussian & \makecell{$0.736 \pm 0.010$ \\ CV=0.014}
             & \makecell{$1.348 \pm 0.006$ \\ CV=0.004}
             & \makecell{$0.971 \pm 0.009$ \\ CV=0.009}
             & \makecell{$2.185 \pm 0.005$ \\ CV=0.002} \\
    Pink     & \makecell{$0.934 \pm 0.010$ \\ CV=0.011}
             & \makecell{$1.479 \pm 0.006$ \\ CV=0.004}
             & \makecell{$1.062 \pm 0.010$ \\ CV=0.009}
             & \makecell{$1.647 \pm 0.037$ \\ CV=0.023} \\
    Brownian & \makecell{$1.100 \pm 0.014$ \\ CV=0.013}
             & \makecell{$1.449 \pm 0.009$ \\ CV=0.006}
             & \makecell{$1.067 \pm 0.013$ \\ CV=0.012}
             & \makecell{$0.048 \pm 0.021$ \\ CV=0.468} \\
    \bottomrule
  \end{tabular}
  \endgroup
\end{table}

These findings confirmed that the ExSEnt method was sensitive to the underlying dynamics of different signals. The ability to capture both the individual and joint complexities of segment durations and amplitudes provided valuable insights into the temporal and amplitude-driven variability of signals. 

\subsection{ExSEnt for representative nonlinear dynamical systems}
\label{subsec: ExSEnt for nonlinear systems}

\noindent We analyzed three low-dimensional nonlinear dynamical systems, the Logistic Map (LM), the Rössler System (LS), and the Rulkov Map (RM), using the ExSEnt measures. By quantifying the distribution and sequencing of extrema, ExSEnt discriminates periodic, quasi-periodic, and chaotic regimes, and highlights transitions such as period-doubling cascades and intermittency. These metrics thus provide complementary sensitivity to changes across parameter space and yield interpretable signatures that deepen our understanding of each system’s dynamics. The control parameters $r$, $c$ and $\sigma$ for LM and RS, RM, respectively, govern the systems' transition from periodic behavior to chaos, making them ideal test cases for evaluating ExSEnt measures. 

\begin{figure} [ht!]
    \centering
    \includegraphics[width=.95\textwidth]{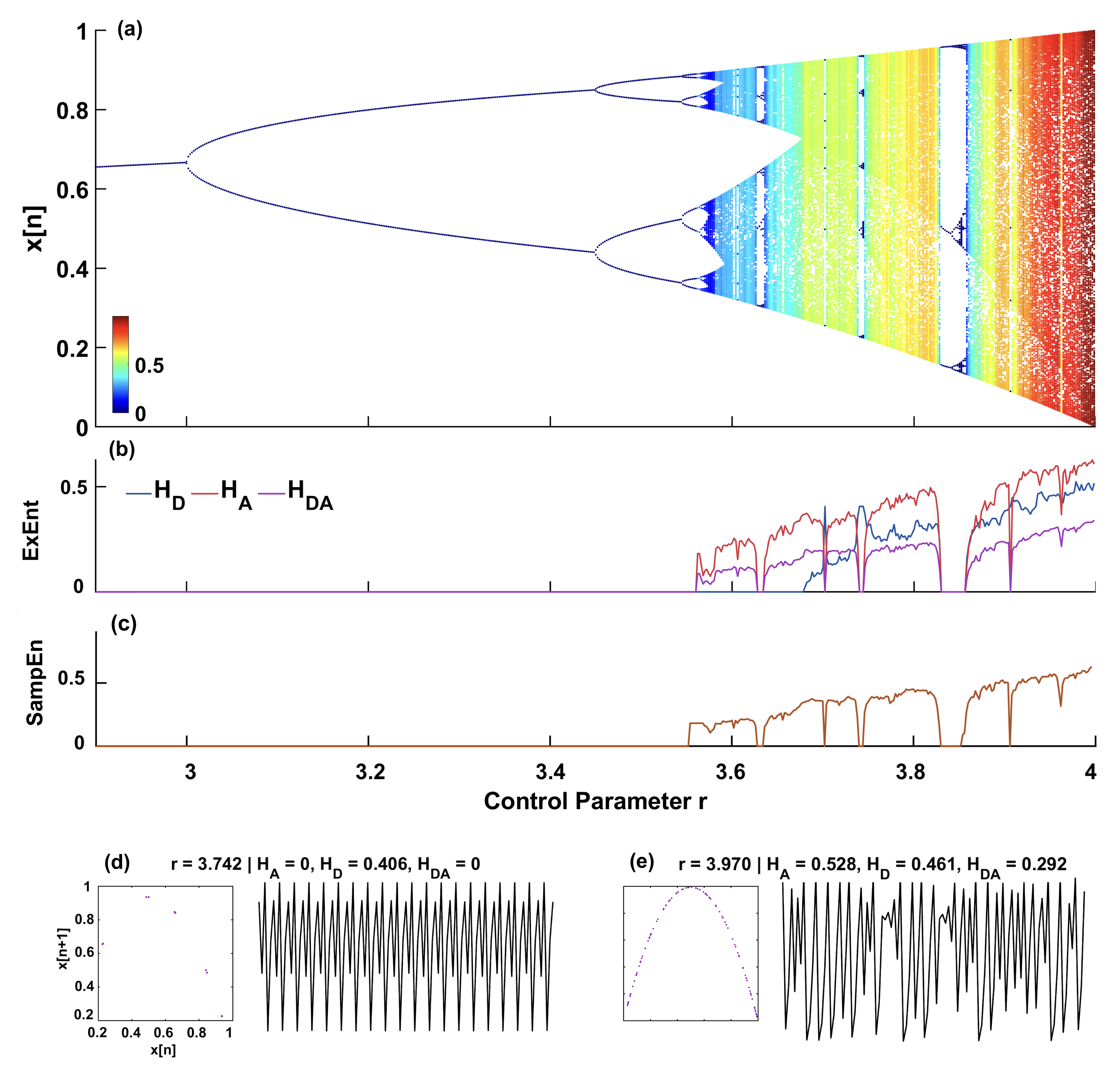}
    \caption{Evaluation of the Logistic map with ExSEnt. (a) Bifurcation diagram of the logistic map, color-coded by normalized joint entropy \( \mathcal{H}_{\mathcal{DA}} \). (b) ExSEnt values for segment duration (\( \mathcal{H}_\mathcal{D} \)), net amplitude (\( \mathcal{H}_\mathcal{A} \)), and joint entropy (\( \mathcal{H}_{\mathcal{DA}} \)); and (c) sample entropy (SampEn), all as a function of the control parameter \( r \). Panels (d) to (g) depict the phase-space and time series of the map for the specified $r$ values along with the corresponding ExSEnt measures. All metrics were computed over \( N=10^4 \) data points.}
    \label{Fig:bifurcation_LM}
\end{figure}

\noindent \emph{Logistic map:} The LM is a one-dimensional discrete-time quadratic map (\ref{subsec: Logistic Map}) that serves as a model of nonlinear behavior and chaos via the period-doubling route \cite{logistic1976simple}. We computed the ExSEnt metrics along with the SampEn, over $r\in[2.9, 4]$. Panel (a) in Fig. \ref{Fig:bifurcation_LM} shows the bifurcation diagram, color-coded based on normalized $\mathcal{H}_\mathcal{DA}$, highlighting regions of varying complexity. Panel (b) shows the ExSEnt measures  $\mathcal{H}_\mathcal{D}$, $\mathcal{H}_\mathcal{A}$, and $\mathcal{H}_\mathcal{DA}$, and panel (c) shows the SampEn metric. While $\mathcal{H}_\mathcal{DA}$ had a similar pattern to SampEn, we observed that $\mathcal{H}_\mathcal{D}$ and $\mathcal{H}_\mathcal{A}$ had distinct values, where $\mathcal{H}_\mathcal{A}$ was larger than $\mathcal{H}_\mathcal{D}$ for most of the range of $r$, except the semi-periodic windows. Panels (d) and (e) show the LM time series and its phase space, with $\mathrm{lag}=1$ for two selected values. For $r=3.742$, while the amplitude and joint entropies were close to zero, the duration entropy was 0.406. When we increased the embedding dimension to $m=3$, the $\mathcal{H}_\mathcal{D}$ was also zero. As we observed, the time series had a period-3 dynamics for this control parameter value, which was different than the embedding dimension of the amplitudes, $m=2$. Panel (e) shows a dynamic where the entropies of amplitude and durations were both higher than their joint entropy.

\begin{figure}[ht!]
\centering
\includegraphics[width=0.9\textwidth]{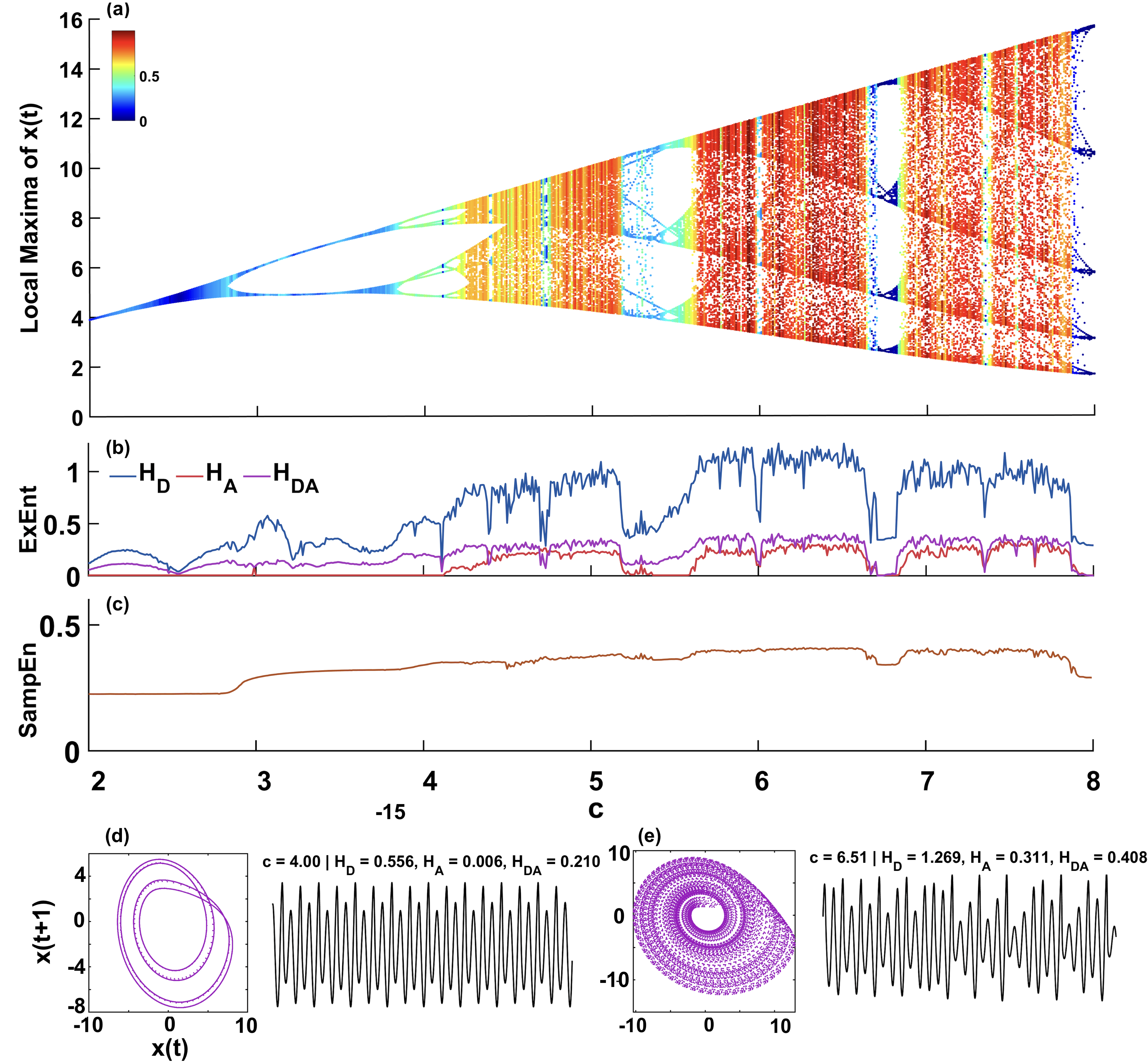}
\vspace*{-2mm}
\caption{Bifurcation diagram and entropy measures of the Rössler system. Panel (a) shows the bifurcation diagram color-coded by the normalized joint entropy $\mathcal{H}_{\mathcal{DA}}$. The ExSEnt metrics $\mathcal{H}_{\mathcal{D}}$, $\mathcal{H}_{\mathcal{A}}$, and $\mathcal{H}_{\mathcal{DA}}$ are depicted as a function of the control parameter $c$ in panel (b). The parameters used were $a=0.2$ and $b=0.2$, with sampling frequency $f_s=10$, $N=10^4$ data points, and $c\in[2,8]$. Sample entropy of the signal over the given range of the control parameter is also shown (panel c). Panels (d) and (e) show the phase space and time series for $c=4.00$ and $c=6.51$, at two different regimes—semi-periodic and semi-chaotic, respectively—with their ExSEnt metrics. The ExSEnt metric provides a measure of information-theoretic complexity for each case. The selected values of $c$ highlight transitions between different dynamical regimes.}
\label{Fig:bifurcation_RS}
\end{figure}

\noindent \emph{Rössler System:} The RS is a simple three-dimensional continuous-time dynamical model that exhibits chaos (\ref{subsec: Rössler System}) and is widely used as a minimalist benchmark in nonlinear dynamics \cite{rossler1976equation}. The bifurcation diagram of the RS, color-coded by the normalized $\mathcal{H}_\mathcal{DA}$, effectively highlights chaotic intervals versus semi-periodic regimes (Fig.~\ref{Fig:bifurcation_RS}, panel a). Among the ExSEnt metrics, the duration-based entropy was higher than the other two (panel b). For semi-periodic parameter values, $\mathcal{H}_{\mathcal{A}}$ and $\mathcal{H}_{\mathcal{DA}}$ were relatively low ($<0.1$), whereas $\mathcal{H}_{\mathcal{D}}$ remained consistently high ($\geq 0.5$) across most of the examined range. Unlike LM, the SampEn for RS did not mirror the joint ExSEnt; instead, it more closely resembled the average of the three ExSEnt measures (panel c). Panels (d) and (e) illustrate two distinct regimes. For $c=4$, $\mathcal{H}_{\mathcal{A}}=0.006$ whereas $\mathcal{H}_{\mathcal{D}}=0.556$. Setting $m=3$ yielded $\mathcal{H}_\mathcal{D}=0.172$ and $\mathcal{H}_{\mathcal{DA}}=0.082$. For $c=6.51$, we had the highest $\mathcal{H}_\mathcal{D}$ value, where the phase space and time series are depicted in panel (e). 

\begin{figure}[ht!]
    \centering
    \includegraphics[width=0.9\textwidth]{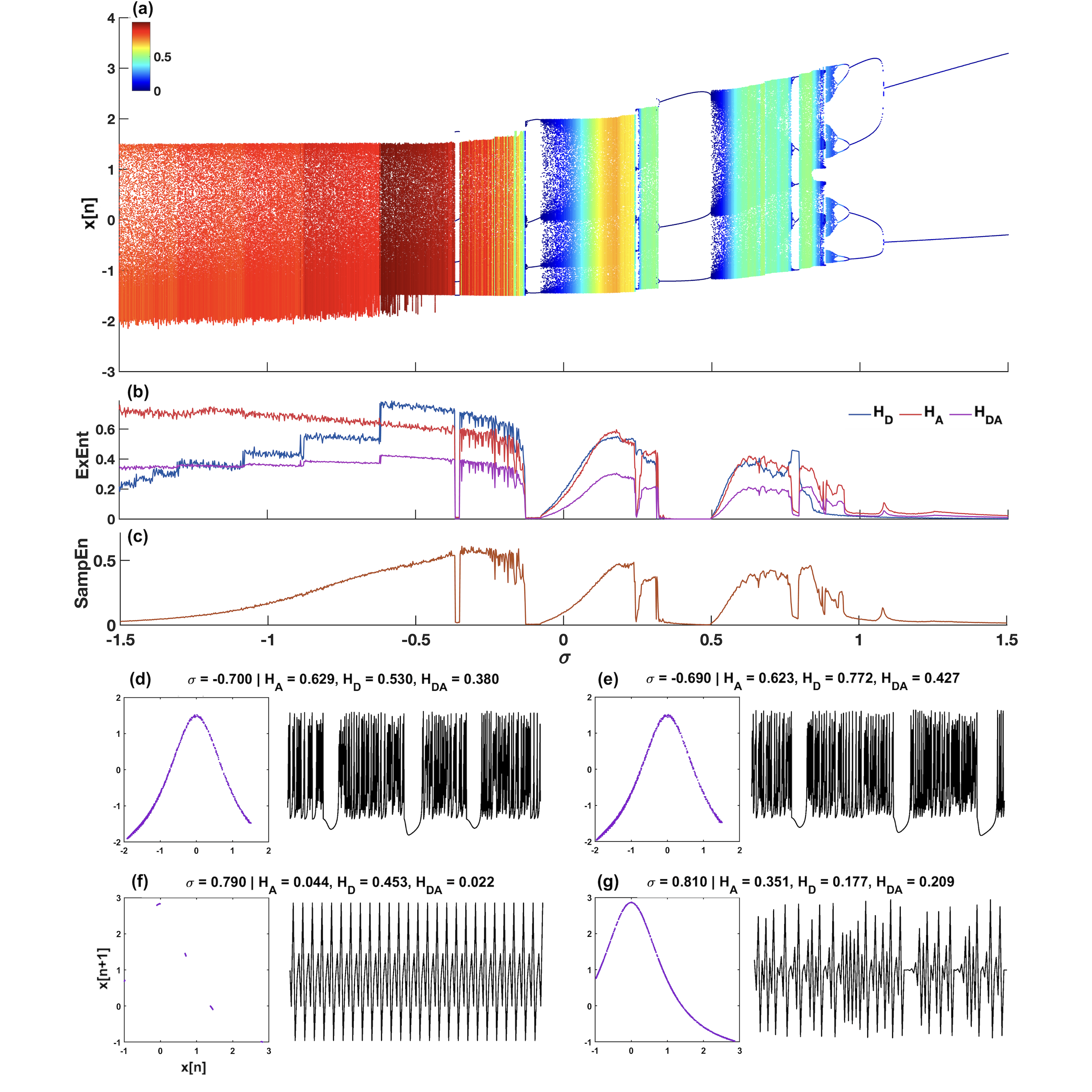}
    \vspace*{-2mm}
    \caption{Bifurcation diagram and entropy measures of the Rulkov map. Panel (a) shows the bifurcation diagram color-coded by the normalized joint entropy $\mathcal{H}_{\mathcal{DA}}$. The ExSEnt metrics as a function of the control parameter $\sigma$ are depicted in panel (b). The parameters used were $\alpha=4.3$, $\mu=0.001$, $N=1.5\times 10^4$ data points, and $\sigma\in[-1.5,1.5]$. Sample entropy over the given range of the control parameter is also shown (panel c). Panels (d)–(g) show the phase space and time series for $\sigma\in\{-0.700,-0.690,0.790,0.810\}$, illustrating different regimes (semi-periodic and semi-chaotic), with their corresponding ExSEnt metrics.}
    \label{Fig:bifurcation_RM}
\end{figure}

\noindent \emph{Rulkov map:} RM is a two-dimensional discrete-time neuron model (\ref{subsec: Rössler System}) that generates key firing patterns, such as silence, tonic spiking, bursting, mixed-mode oscillations, and chaos, via slow–fast dynamics \cite{rulkov2002modeling}.  Because it’s computationally lightweight yet physiologically interpretable, it is widely used for large-scale network simulations and for bifurcation-based studies of excitability and bursting regimes. The bifurcation diagram of the RM for control parameter $\sigma$, color-coded by the joint ExSEnt, highlights chaotic bursting intervals versus semi-periodic regimes (Fig.~\ref{Fig:bifurcation_RM}, panel a). Comparing panels (b) and (c), although the overall pattern of SampEn resembled the smoothed $\mathcal{H}_{\mathcal{D}}$ and $\mathcal{H}_{\mathcal{DA}}$ (panel c), the ExSEnt measures exhibited richer variability (panel b). While $\mathcal{H}_{\mathcal{A}}$ was relatively smooth for $\sigma < -0.368$, $\mathcal{H}_{\mathcal{D}}$ and $\mathcal{H}_{\mathcal{DA}}$ showed step-like rises at $\sigma \in \{-1.076,-0.878,-0.620\}$, whereas SampEn remained smooth at these values. Another noteworthy interval was $\sigma \in [0.768,0.796]$, where $\mathcal{H}_{\mathcal{A}}$, $\mathcal{H}_{\mathcal{DA}}$, and SampEn were near zero ($\le 0.07$), while $\mathcal{H}_{\mathcal{D}} \ge 0.4$. Panels (d) and (e) focus on two parameter values within this step-like region: although the time series and phase portraits for $\sigma=-0.700$ and $\sigma=-0.690$ appeared similar, and the entropies of $\mathcal{A}$ and $\mathcal{DA}$ were comparable, we observe a $44.66\%$ increase in the entropy of $\mathcal{D}$ in the latter case compared to the former. We also inspected $\sigma=0.790$ (semi-periodic; panel f) and $\sigma=0.810$ (semi-chaotic), whose phase spaces corroborated these classifications. The $\mathcal{H}_\mathcal{D}=0.453$, for $\sigma=0.790$ for $m=2$ was high, which raised the question of whether the effective embedding dimension of $\mathcal{D}$ was higher than 2. Increasing the embedding dimension to $m=5$ yielded $\mathcal{H}_{\mathcal{D}}=0.048$, again underscoring the distinct effective embedding of $\mathcal{D}$ relative to the other two metrics.


\begin{figure}[ht!]
    \centering
    \includegraphics[width=0.9\textwidth]{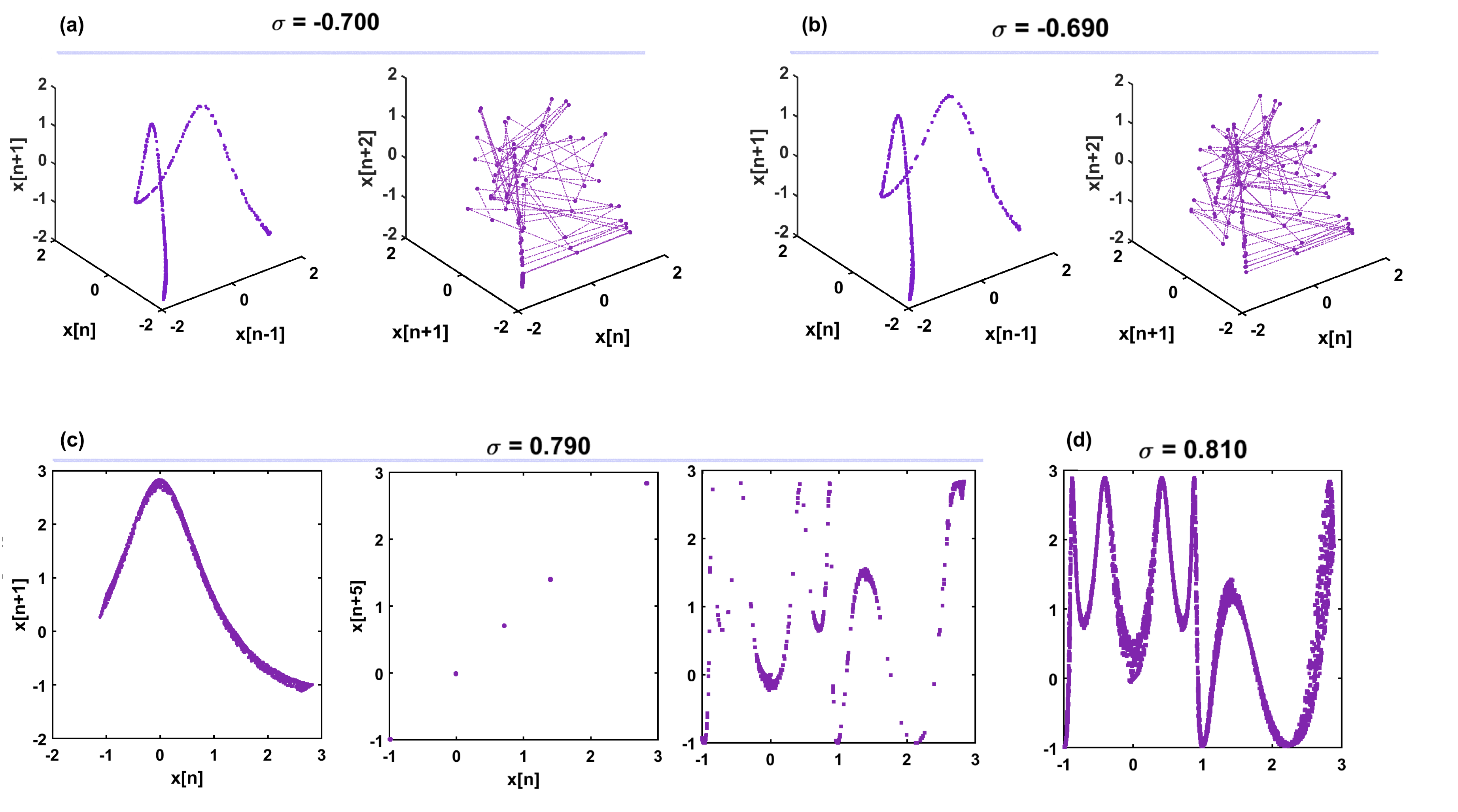}
    \vspace*{-2mm}
    \caption{Phase space for selected $\sigma$-values of the Rulkov map. All the parameters are the same as Fig. \ref{Fig:bifurcation_RM}. Panel (a) and (b) show the phase space of RM for $\sigma\in\{-0.700, -0.690\}$, with lag, $\tau=1$. The left plots show the phase space over $N=5000$ points, and the right plots over $N=50$ points. Panel (c) shows the phase space of RM for $\sigma=0.790$. The first plot on the left shows the phase space over $N=2\times10^4$ data points. The middle panel shows the phase space with $\tau=5$ for $N=500$, and the plot on the right shows the same phase space over $N=5000$. Panel (d) shows the phase space of RM for $\sigma=0.810$ with $\tau=5$, over $N=5000$. } 
    \label{Fig:RM_samples} 
\end{figure}

Fig.~\ref{Fig:RM_samples} provides a closer look at the $\sigma=-0.700$ and $-0.690$ phase spaces. Over long segments, both parameters show similar patterns at the three-dimensional lag one phase-space  (panels a and b, left plots). However, when we examined a limited segment and plotted the phase space with lines connecting consecutive states, $\sigma=0.690$ exhibited a more complex traversal between states than $\sigma=0.700$. This is consistent with the higher $\mathcal{H}_{\mathcal{D}}$ reported earlier. Hence, the separate calculation of the entropy of durations and amplitude of segments gives a better perspective of the signal's trajectory and dynamics.

Panel (c) in Fig. \ref{Fig:RM_samples} displays three representations of the phase space for the RM system at $\sigma=0.790$. The left plot illustrates the phase space over an extended range ($N=10^4$), exhibiting similarity to the $\tau=1$ phase space observed at $\sigma=0.810$ (Fig. \ref{Fig:bifurcation_RM}, panel g). However, the $\mathcal{H}_\mathcal{D}$ metric yielded a higher value in the former case, which appeared inconsistent with the visual characteristics of the phase space. To better understand the underlying dynamics, we examined the phase space at different lags. Specifically for $\tau=5$, and $N=500$, phase space revealed a periodic pattern with $\mathcal{H}_\mathcal{D}=0.084$, whereas analysis of a longer data range ($N=10^4$) demonstrated that the system manifests a complex pattern for this control parameter. This indicates that the semi-periodic behavior occurs only over limited intervals, while the system transitions toward more chaotic phases at extended scales. Panel (d) shows the phase space at $\tau=5$ for $\sigma=0.810$, which resembles the lag-5 structure of the $\sigma=0.790$ case but exhibits greater state density due to its more continuous appearance, accompanied by $\mathcal{H}_\mathcal{D}=0.192$.

These findings highlight the capability of ExSEnt to distinguish different dynamical regimes and capture subtle complexity variations that are not fully resolved by conventional entropy measures. By effectively characterizing both individual and joint entropy contributions, ExSEnt provides a nuanced understanding of the interplay between temporal and amplitude fluctuations across different system states. Various methods and algorithms are proposed to determine the optimal embedding delay and dimension, such as the use of mutual information for estimating time delays \cite{fraser1986using} and different methods for embedding dimension calculation such as the false nearest neighbors approach \cite{kennel1992determining}, or Cao’s method, which evaluates the independence of reconstructed vectors in the phase space to identify the minimum sufficient dimension \cite{cao1997practical}. Our observations suggest that when temporal and amplitude-driven patterns of a signal are analyzed separately, they may exhibit distinct characteristics across different lags and embedding dimensions. This implies that the dimensionality of a signal, when viewed through the lens of temporal versus amplitude components, may not necessarily coincide. While a detailed exploration of this perspective lies beyond the scope of the present study, such an approach to signal complexity could offer new insights into the dimensionality and phase space representation of complex systems.

\subsection{Convergence properties of ExEnt measures}
\label{subsec: ExSEnt stability test}

\begin{figure}[ht!] 
    \centering
    \includegraphics[width=1\textwidth]{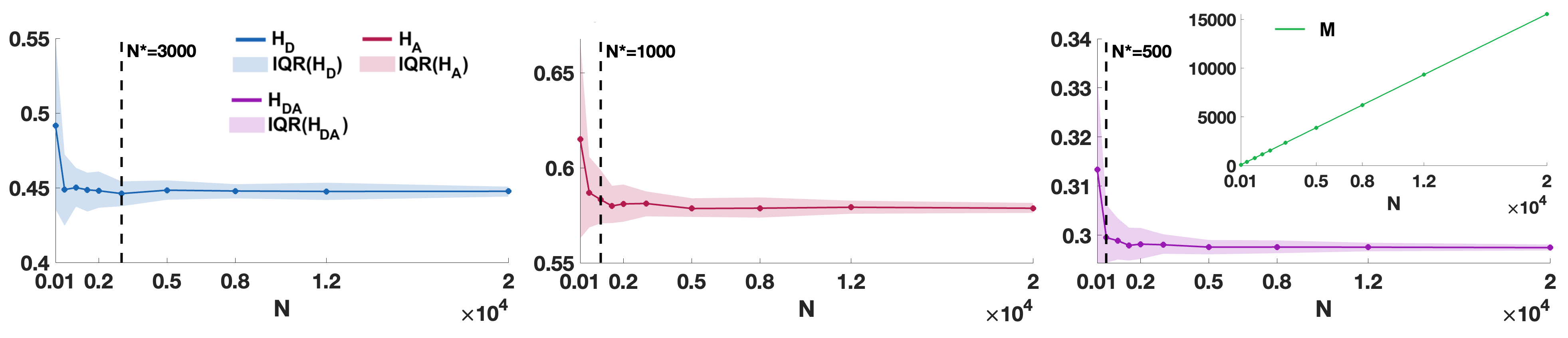}
    \vspace*{-2mm}
    \caption{ExSEnt convergence properties over $N$ on the logistic map. For each data length $N$, the metrics were computed over 100 realizations with random initial conditions. The mean is plotted with a shaded Interquartile Range (IQR) band. The legend identifies each ExSEnt metric. Control parameter was $r=3.95$ and ExSEnt parameters were $\lambda=0.01$. The inset (green curve) shows the number of extracted segments, $M$, versus $N$. A dashed vertical line marks the $\pm2.5\%$ IQR-based stability threshold.}
    \label{fig: ExSEnt stability}
\end{figure}

\noindent The convergence of a metric with respect to data length is critically important, as it ensures consistent characterization of dynamical systems across varying observational scales and prevents misinterpretation of emergent features that may be finite-size effects. We investigated the convergence of ExSEnt metrics as a function of $N$ using the logistic map. Because the number of extracted segments $M$ determines the lengths of the $\mathcal{D}$ and $\mathcal{A}$ sequences, we also quantified how $M$ scales with $N$. The logistic map provides an ideal testbed for convergence analysis due to its measure-preserving properties in the chaotic regime, where invariant statistics persist across distinct trajectories generated by varying initial conditions. We fixed $r=3.95$ and computed the ExSEnt metrics across 100 realizations per $N$ using random initial conditions, with the mean stability curve and its interquartile range (IQR) depicted as a shaded region in Fig. \ref{fig: ExSEnt stability}.

As $N$ increased, the entropy measures converged to stable values, indicating that the method yielded consistent estimates for sufficiently long time series. A dashed black vertical line indicated the smallest $N$ at which the metric entered an acceptable stability range, defined as an interquartile range (IQR) band whose half-width did not exceed $2.5\%$ of the metric’s final value. Under this criterion, stability was reached at $N=3000$ for $\mathcal{H}_{\mathcal{D}}$, at $N=1000$ for $\mathcal{H}_{\mathcal{A}}$ and at $N=500$ for $\mathcal{H}_{\mathcal{DA}}$. The inset shows the mean number of extracted segments as a function of $N$, which increased approximately linearly.

Although $N\approx 3000$ was a reasonable series length for the LM, the relationship between the sample size $N$ and the number of extracted segments, $M$, determines the optimal length, which varies with the signal’s characteristics. For the LM, the $M$ vs. $N$ curve yielded $M\approx0.77\times N$. As expected, $M$ depended on the signal’s frequency content; therefore, the optimal $N$ depended on the data characteristics and should be examined for each dataset.

\subsection{ExSEnt for biosignals}
\label{subsec: ExSEnt for biological data}

\noindent Nonlinear metrics play a critical role in evaluating biosignals because they quantify dynamical properties, such as complexity and irregularity, that linear statistics cannot capture. In particular, entropy-based measures summarize uncertainty and temporal organization and can serve as candidate biomarkers when standardized across cohorts and conditions, being sensitive to pathology-related alterations while robust to moderate noise and amplitude scaling. We extracted ExSEnt measures from two real datasets: surface EMG recordings and inertial measurements.

\begin{figure} [H]
    \centering
    \includegraphics[width=.9\textwidth]{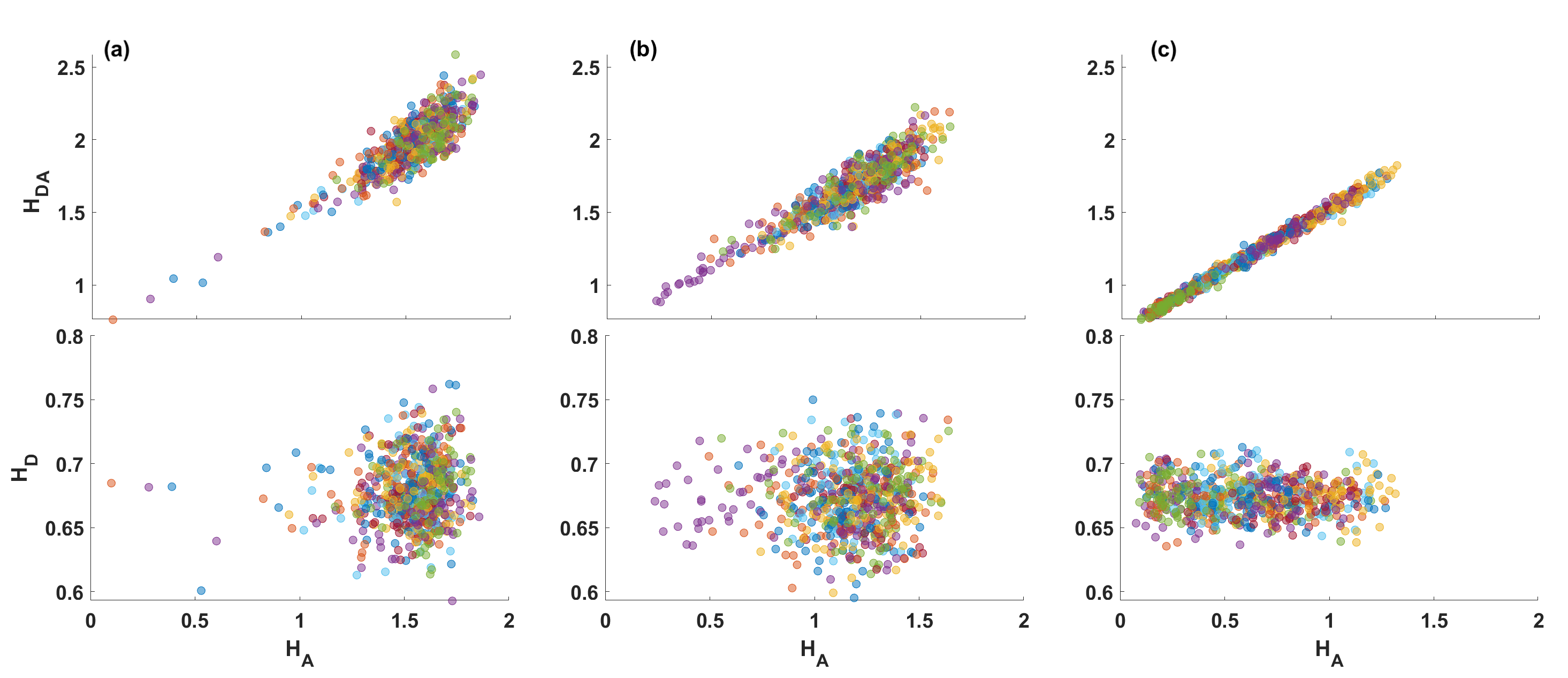}
    \caption{ExSEnt metrics over surface EMG during a finger pinching task: movement reduces amplitude complexity but preserves temporal entropy. Panel (a), top, shows joint ExSEnt ($\mathcal{H}_{\mathcal{DA}}$) vs. amplitude entropy ($\mathcal{H}_\mathcal{A}$) during a 2\,s baseline interval; the bottom scatter plot shows duration entropy ($\mathcal{H}_\mathcal{D}$) vs. $\mathcal{H}_\mathcal{A}$ for the same interval. Panel (b), top and bottom, is analogous to panel (a) but for a 2\,s window extracted from the movement period. Panel (c), top and bottom, shows the same metrics over the entire 7\,s task window. All metrics are computed over the 100--200\,Hz band. }
    \label{Fig: EMG ExSEnt}
\end{figure}

\noindent \emph{EMG analysis:} We applied ExSEnt to EMG signals recorded from the forearms of $19$ healthy subjects during a finger-pinching task. We evaluated the metrics in three movement phases: (1) a 2-second baseline interval, during which the subjects rested their arms on the table while awaiting the go-cue command; (2) a 2-second window during the execution of the movement; and (3) the entire 7-second trial (Fig.~\ref{Fig: EMG ExSEnt}). We observed that the movement decreased $\mathcal{H}_\mathcal{A}$ considerably ($-23.95\%$), while $\mathcal{H}_\mathcal{D}$ only showed a slight reduction ($-0.92\%$), indicating regularized amplitude modulation with substantially smaller decline in temporal complexity. The data and the preprocessing steps were described in \cite{Cho17, kamali2024mu}.

The EMG signals during the motor task revealed that movement intervals exhibit lower amplitude complexity but relatively stable temporal entropy compared to the baseline. A strong linear correlation was observed between the $\mathcal{H}_\mathcal{A}$ and $\mathcal{H}_\mathcal{DA}$, suggesting that amplitude variations are the primary contributors to the joint amplitude-temporal EMG complexity. The complexity measures displayed the most prominent difference in the $100–200$\,Hz frequency range, in 7-second intervals, aligning with known motor activity dynamics.

\noindent \emph{Ankle motion acceleration} The Daphnet Freezing of Gait dataset documents a wearable assistant for Parkinson’s disease that uses on-body accelerometers to monitor movement, detect freezing of gait (FOG), and deliver rhythmic auditory cues to help patients resume walking \cite{Daphnet}. Ten PD patients performed laboratory walking tasks; eight patients exhibited FOG, and FOG events were verified from synchronized video recordings \cite{Daphnet}.

\begin{table}[ht]
\centering
\begingroup
\setlength{\tabcolsep}{4pt}
\renewcommand{\arraystretch}{0.95}
\scriptsize
\makebox[\textwidth]{ 
\begin{tabular}{@{}lcc@{}} 
\toprule
Metric & 0.5--3\,Hz & 3--8\,Hz \\
\midrule
HD  & --            & \checkmark \\
HA  & \checkmark    & \checkmark \\
HDA & --            & \checkmark \\
\bottomrule
\end{tabular}
}
\endgroup
\caption{Wilcoxon rank-sum significance by band (two-sided, $p=0.05$) for ExSEnt metrics over VAA of Daphnet dataset.}
\label{tab:ranksum_daphnet}
\end{table}

\begin{figure} [ht!]
    \centering
    \includegraphics[width=1\textwidth]{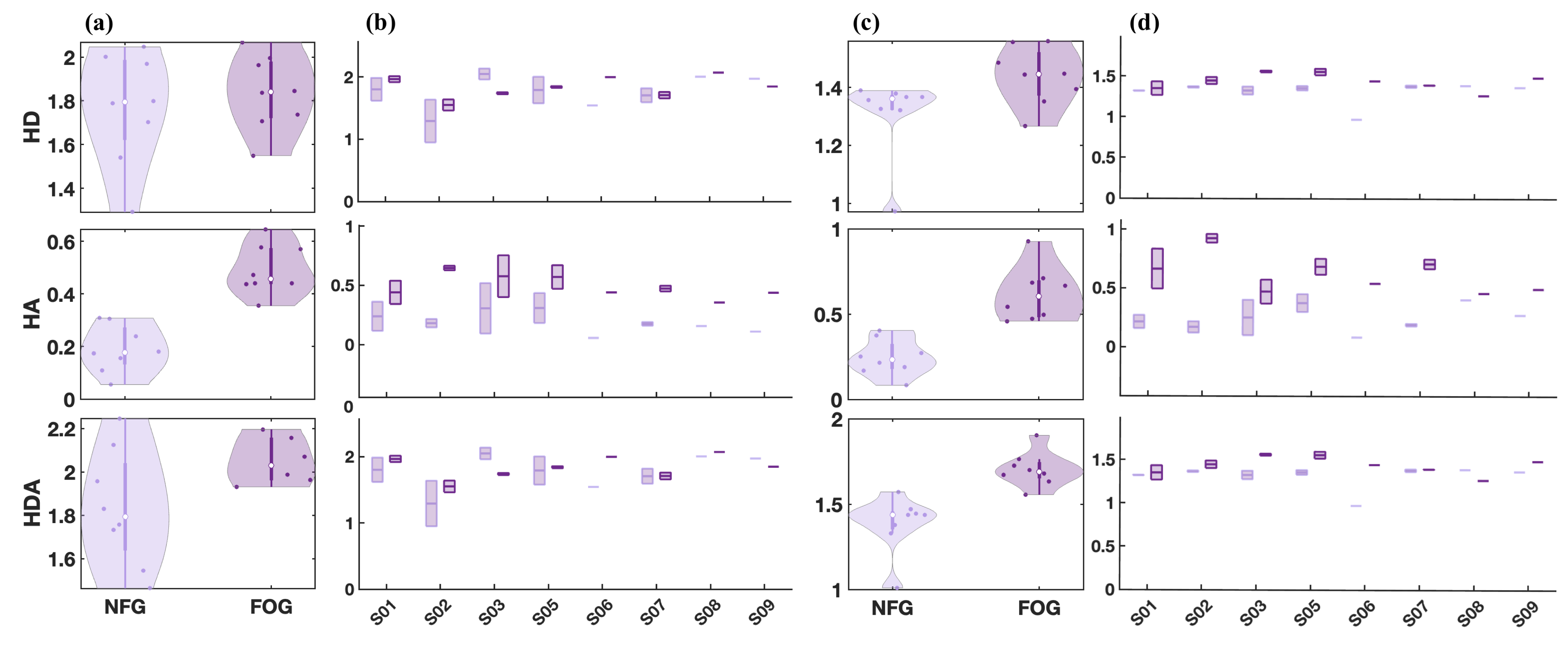}
    \caption{ExSEnt metrics of vertical ankle acceleration (VAA) in Parkinson’s disease. Panels (a) and (b) show the 0.5--3\,Hz band and (c) and (d) the 3--8\,Hz band of the band-pass filtered VAA for all three metrics. Panels (a) and (c) show the violin plots comparing no-freezing gait (NFG) and freezing of gait (FOG) using the subject-level mean of each metric. Panels (b) and (d) depict the subject-wise boxplots for NFG and FOG, for the ExSEnt metrics, showing per-subject distributions.}
    \label{Fig: Daphnet violin}
\end{figure}

We analyzed vertical ankle acceleration (VAA) using the standard locomotor (0.5--3\,Hz) and freeze (3--8\,Hz) bands; the Freeze Index (FI)—the 3--8\,Hz/0.5--3\,Hz power ratio—has been validated with ankle/shank sensors in lab and ambulatory settings \cite{moore2008ambulatory,moore2013autonomous}. We computed the three ExSEnt metrics ($\mathcal{H}_\mathcal{A}$, $\mathcal{H}_\mathcal{D}$, $\mathcal{H}_\mathcal{DA}$) in both bands for NFG and FOG windows. NFG windows contained no freezing episodes, whereas FOG windows comprised the majority ($\geq 70\%$) of their duration in the freezing state. All windows were at least 30\,s long. ExSEnt differed significantly between NFG and FOG for $\mathcal{H}_\mathcal{A}$ in the 0.5--3\,Hz band and for all three metrics in the 3--8\,Hz band (Table~\ref{tab:ranksum_daphnet}). For comparison, SampEn on the same signals also discriminated NFG from FOG (Fig.~\ref{Fig: Daphnet violin}); however, ExSEnt resolved band-specific structure: in the 0.5--3\,Hz band only amplitude entropy ($\mathcal{H}_\mathcal{A}$) differed significantly, whereas in the 3--8\,Hz band all three metrics ($\mathcal{H}_\mathcal{A}$, $\mathcal{H}_\mathcal{D}$, $\mathcal{H}_\mathcal{DA}$) were altered. Violin plots of subject-level averages show non-overlapping $\mathcal{H}_\mathcal{A}$ distributions in both bands (panels a and c), and subject-wise boxplots further illustrate ExSEnt’s ability to distinguish gait from freezing (panels b and d).

\begin{figure} [ht!]
    \centering
    \includegraphics[width=1\textwidth]{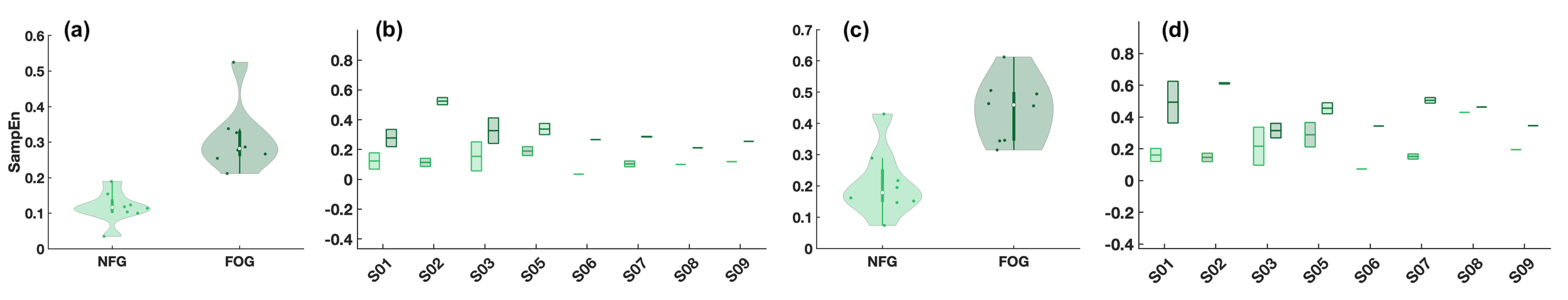}
    \caption{Sample entropy of vertical ankle acceleration (VAA) in Parkinson’s disease. Panels (a) and (b) show the 0.5--3\,Hz band and (c) and (d) the 3--8\,Hz band of the band-pass filtered VAA. Panels (a) and (c) show the violin plots comparing no-freezing gait (NFG) and freezing of gait (FOG) using the subject-level mean of each metric. Panels (b) and (d) depict the subject-wise boxplots for NFG and FOG, showing per-subject distributions.}
    \label{Fig: Daphnet sampen}
\end{figure}

\section{Discussion}
\label{sec: Discussion}

\noindent The ExSEnt metrics provides deeper insight into the structural organization of nonlinear dynamical systems by distinguishing the source of entropy—whether it arises primarily from temporal irregularity (segment durations) or from amplitude variability (segment-wise net changes). Across bifurcation diagrams, the three indices ($\mathcal{H}_\mathcal{D}$, $\mathcal{H}_\mathcal{A}$, and the joint $\mathcal{H}_{\mathcal{DA}}$) offer complementary perspectives on the underlying dynamics. Notably, the ExSEnt indices do not simply co-vary; instead, they differentiate temporal from amplitude-driven contributions to complexity, revealing structure that a single, undifferentiated complexity score would confuse.

In semi-periodic regions of the bifurcation diagrams (Figs.~\ref{Fig:bifurcation_LM}--\ref{Fig:bifurcation_RM}), intervals were observed in which either $\mathcal{H}_\mathcal{D}$ or $\mathcal{H}_\mathcal{A}$ dominated the joint behavior, and $\mathcal{H}_{\mathcal{DA}}$ closely tracked the dominant component. This indicated that one feature (timing or amplitude) was the primary driver of system complexity in those parameter ranges. Such feature-specific dominance was not captured by traditional complexity measures that conflate irregularities in timing and amplitude. By isolating the contributions of durations and amplitudes, ExSEnt yielded a more nuanced characterization of the trajectory, exposing patterns in the temporal organization or in the amplitude modulation that conventional entropy metrics overlooked.

There were also regions of the bifurcation diagrams (e.g., Fig.~\ref{Fig:bifurcation_RM}, $0.5<\sigma<0.7$) where the trajectory exhibited a high density of extrema while the ExSEnt indices remained comparatively low relative to other ranges (e.g., $-1.5<\sigma<0.4$). This dissociation suggested that frequent extrema alone did not imply high complexity; rather, the distribution of extrema followed a more regular pattern, reducing entropy. In that referred region, both $\mathcal{H}_\mathcal{D}$ and $\mathcal{H}_\mathcal{A}$ exceeded the joint entropy $\mathcal{H}_{\mathcal{DA}}$, implying partial redundancy between temporal and amplitude variations (i.e., some predictability of one feature given the other).

On representative biological data, ExSEnt provided interpretable, feature-level insights. In the EMG recordings, the amplitude-driven entropy $\mathcal{H}_\mathcal{A}$ decreased during movement execution relative to baseline, while the temporal component $\mathcal{H}_\mathcal{D}$ remained approximately unchanged. Although EMG amplitudes increased when muscles were engaged, the pattern of amplitude changes at extrema became \emph{less} complex than at rest; this was reflected proportionally in the joint index $\mathcal{H}_{\mathcal{DA}}$. Thus, ExSEnt separated a power/gain increase from a reduction in amplitude \emph{complexity}, which a single-value entropy would have conflated.

In VAA from participants with PD, ExSEnt cleanly separated band-specific effects of FOG. In the 3--8\,Hz band, all three indices ($\mathcal{H}_\mathcal{D}$, $\mathcal{H}_\mathcal{A}$, $\mathcal{H}_{\mathcal{DA}}$) increased, indicating concurrent rises in temporal and amplitude irregularities. In contrast, within 0.5--3\,Hz only $\mathcal{H}_\mathcal{A}$ increased, implicating amplitude-driven changes at slow time scales while timing irregularity ($\mathcal{H}_\mathcal{D}$) remained stable. This accords with prior FI-based evidence that FOG is characterized by enhanced 3--8\,Hz activity with comparatively stable locomotor band content \cite{moore2013autonomous,moore2008ambulatory}. ExSEnt makes the attribution explicit by showing that, in the locomotor band, durations preserve their entropy while amplitudes exhibit increased complexity.

Overall, ExSEnt effectively characterized complexity variations in both canonical dynamical systems and real-world physiological signals. The joint entropy $\mathcal{H}_{\mathcal{DA}}$ provided a clear quantitative index that was consistent with established entropy-based assessments, while the separate components identified the \emph{primary contributor} to complexity (temporal vs.\ amplitude). In some cases (e.g., Fig.~\ref{Fig:RM_samples}), the results indicated differing dimensionalities for temporal and amplitude-driven dynamics, reinforcing that these features need not vary on the same manifold. Peaks and drops in ExSEnt measures aligned with transitions between periodic and chaotic behavior, capturing critical dynamical shifts. Importantly, incorporating a noise-tolerance threshold via the tunable parameter $\lambda$ improved robustness to small perturbations and observational noise, enhancing suitability for real, noisy systems.

In summary, ExSEnt provides a feature-driven perspective on signal complexity that complements traditional measures. By attributing overall entropy to timing and amplitude components and quantifying their interaction, ExSEnt improves interpretability and offers practical advantages for analyzing nonlinear dynamics in both synthetic and real data.

\section{Conclusion}
\label{sec:conclusion}

We introduced ExSEnt, a feature-decomposed complexity metric that characterizes the dynamics of nonlinear signals by separating temporal from amplitude contributions. The method segments a signal into monotonic portions defined by the sign of first-order increments, with an IQR-based threshold to provide robustness against noise-driven fluctuations. From each segment, we extract the duration $\mathcal{D}$ and the net amplitude change $\mathcal{A}$, then compute sample entropy on the duration sequence ($\mathcal{H}_{\mathcal D}$), the amplitude sequence ($\mathcal{H}_{\mathcal A}$), and their joint representation ($\mathcal{H}_{\mathcal{DA}}$). Using SampEn is advantageous after segmentation because the effective sample size is reduced.

Comparing these indices yields interpretable insight into \emph{where} complexity originates. High $\mathcal{H}_{\mathcal D}$ indicates irregular timing of salient events, whereas high $\mathcal{H}_{\mathcal A}$ reflects irregularity in amplitude excursions. The joint term $\mathcal{H}_{\mathcal{DA}}$ quantifies coupling: values lower than both individual components suggest redundancy (predictability of one feature from the other), while similarly high values across all three indicate independent temporal and amplitude contributions to overall complexity. Across canonical systems, we observed parameter ranges in which temporal and amplitude components follow divergent regimes, underscoring that feature-specific dynamics need not share the same effective dimensionality.

Empirically, ExSEnt tracked complexity changes along control parameter sweeps in the Logistic map, R\"ossler system, and Rulkov map, aligning with transitions between periodic and chaotic behavior. In biosignals, ExSEnt isolated feature-level effects (e.g., band-specific changes in ankle acceleration during FOG vs.\ NFG, and amplitude-complexity reduction in EMG during movement despite power increases), providing a richer perspective than unidimensional entropy measures. The increment-based segmentation and noise-tolerance threshold improved robustness to small perturbations, making ExSEnt suitable for real-world, noisy recordings.

ExSEnt has broad applicability wherever disentangling timing and amplitude variability is informative, notably in physiological and biosignal analysis (EEG/MEG, EMG, gait and wearables, HRV, sleep and anesthesia monitoring, seizure detection), as well as in finance, climate and environmental monitoring, and condition-based maintenance of engineered systems. Future work includes multiscale extensions (to attribute feature-specific complexity across time scales), principled parameter selection and theoretical bounds, multivariate/graph extensions for networked signals, the use of application-specific relevant events instead of extrema, and real-time implementations for embedded and clinical settings. Furthermore, ExSEnt metrics can be used together with state-of-the-art AI algorithms for time series classification. 

In sum, ExSEnt complements traditional entropy methods by attributing complexity to distinct signal features and quantifying their interaction, thereby improving interpretability and advancing the analysis of complex dynamical systems and signals.

\section*{Code Availability}
The implementation of the ExSEnt method is available at:
\href{https://github.com/GNB-UAM/Kamali_et_al_ExSEnt_Extrema-segmented_Entropy_Analysis_of_Time_Series}{GitHub.com/GNB-UAM/ExSEnt-Analysis}.

\appendix
\section{Mathematical Models}
\label{app1}

\subsection{Stochastic Signal Generation}
\label{math sec: stochastic}
To validate the entropy-based metrics, we generated three canonical stochastic signals over $N=2\times10^4$ samples with a sampling rate of \(f_s=200\) Hz. Each signal was simulated 100 times with fixed random seeds to ensure reproducibility. The statistical descriptors reported in Table \ref{tab:exsent_results} were computed across runs. Below, we summarize the mathematical formulation of these signals.

\subsubsection{Gaussian White Noise}
Gaussian white noise was generated as samples from the standard normal distribution, $\mathcal{N}(0,1)$, whose probability density function is
\begin{equation}
p(x) = \frac{1}{\sqrt{2\pi\sigma^2}} 
       \exp\!\left(-\frac{(x-\mu)^2}{2\sigma^2}\right),
\end{equation}
with mean $\mu = 0 $ and standard deviation $\sigma = 1$, generated with the MATLAB command \texttt{rand(N,1)}.

\subsubsection{Pink Noise}
Pink noise ($1/f$ noise) was generated as a stochastic process with a power spectral density inversely proportional to frequency:
\begin{equation}
S(f) \propto \frac{1}{f}, \quad f > 0,
\end{equation}
which yields equal energy per octave. In practice, it was generated using MATLAB's 
\texttt{dsp.ColoredNoise('Color','pink',N,1)} function. 

\subsubsection{Brownian Motion}
Brownian motion was generated as the cumulative sum of independent Gaussian increments:
\begin{equation}
x[n] = \sum_{k=1}^{n} \eta[k], \qquad \eta[k] \sim \mathcal{N}(0,1).
\end{equation}
This corresponds to the discrete-time equivalent of integrated white noise, with a power spectral density scaling as
\begin{equation}
S(f) \propto \frac{1}{f^2}.
\end{equation}
In MATLAB, it was implemented as \texttt{cumsum(randn(N,1))}.

\subsection{Logistic Map}
\label{subsec: Logistic Map}
\noindent The logistic map is a simple nonlinear dynamical system defined by the recurrence relation:
\begin{equation}
    x_{n+1} = r x_n (1 - x_n),
\end{equation}
where \( x_n \in [0,1] \) represents the population at iteration \( n \), and \( r \) is the control parameter governing the system's behavior. For \( 0 < r \leq 4 \), the map exhibits a range of dynamical behaviors, including fixed points, periodic cycles, and chaos.

\subsection{Rössler System}
\label{subsec: Rössler System}
\noindent The Rössler system is a set of three coupled first-order autonomous differential equations that describe a continuous-time chaotic attractor:
\begin{align}
    \frac{dx}{dt} &= - y - z, \\
    \frac{dy}{dt} &= x + a y, \\
    \frac{dz}{dt} &= b + z (x - c),
\end{align}
where \( a, b, c \) are system parameters that influence the system’s behavior. In particular, \( c \) serves as a control parameter, impacting the transition between periodic and chaotic dynamics. A commonly studied set of parameters is \( (a, b, c) = (0.2, 0.2, 5.7) \), which produces chaotic dynamics.

\subsection{The Rulkov Map}
\label{subsec: The Rulkov Map}
\noindent The Rulkov map is a two-dimensional discrete-time system designed to model neuron-like bursting dynamics. It consists of a fast variable \( x_n \), representing the membrane potential, and a slow variable \( y_n \), which modulates the system’s state over time:
\begin{equation}
    x_{n+1} = \frac{\alpha}{1 + x_n^2} + y_n,
\end{equation}
\begin{equation}
    y_{n+1} = y_n - \mu (x_n - \sigma).
\end{equation}
The parameter \( \alpha \) primarily controls the excitability of the system and typically ranges from \( 2 \) to \( 6 \), where lower values correspond to tonic spiking and higher values induce chaotic bursting. The small parameter \( \mu \), governing the slow evolution of \( y_n \), is usually set within \( 10^{-4} \leq \mu \leq 10^{-2} \), ensuring a clear timescale separation between fast and slow dynamics. The parameter \( \sigma \) acts as an external input or bifurcation parameter, influencing transitions between different dynamical regimes, with common values in the range \( -1 \leq \sigma \leq 1 \). The interplay between these parameters allows the Rulkov map to generate a variety of neuronal activity patterns, including periodic spiking, bursting, and chaotic oscillations.

\section*{Acknowledgments}

This research was funded by grants 
PID2024-155923NB-I00, CPP2023-010818, and 
PID2021-122347NB-I00 (MCIN/AEI and ERDF- "A way of making Europe").

\bibliographystyle{sn-mathphys}
\bibliography{references}

\end{document}